\documentclass[preprint,12pt]{elsarticle}

\usepackage{amssymb}
\usepackage{amsmath}

\usepackage{lineno}
\usepackage{graphicx}
\usepackage{dcolumn}
\usepackage{bm}
\usepackage{xcolor}
\usepackage{adjustbox}
\usepackage[utf8]{inputenc}
\usepackage[T1]{fontenc}
\usepackage{natbib}
\usepackage[colorlinks=true, linkcolor=blue, citecolor=blue, urlcolor=blue]{hyperref}

\journal{Physics of the Dark Universe}

\begin{document}

\begin{frontmatter}



\title{\boldmath Revisiting $\Lambda$CDM extensions in light of re-analyzed CMB data} 


\author[label1]{Jacobo Asorey\corref{cor1}}
\author[label2]{Javier de Cruz P\'erez\corref{cor2}}
\cortext[cor1]{jasorey@unizar.es}
\cortext[cor2]{jdecruz@uco.es}
\affiliation[label1]{organization={Departamento de F\'{\i}sica Te\'orica, Centro de Astropart\'iculas y F\'isica de Altas Energ\'ias (CAPA), Universidad de Zaragoza},
            city={Zaragoza},
            postcode={50009}, 
            country={Spain}}
\affiliation[label2]{organization={Departamento de F\'{\i}sica, Universidad de C\'ordoba},
            addressline={Campus Universitario de Rabanales, Ctra. N-IV km, 396}, 
            city={C\'ordoba},
            postcode={14071}, 
            country={Spain}}            

\begin{abstract}
In the last years with the increasing precision in cosmological observations we have been able to establish a standard model of cosmology, the so-called $\Lambda$CDM, but also find some tensions between cosmological probes that are difficult to explain within the context of this model. We tested several phenomenological extensions of the $\Lambda$CDM with the newest datasets from the chain CMB+BAO+SNIa, to see whether they are able to alleviate the aforementioned tensions. We find that when the updated version of the Planck CMB likelihood (PR4  \texttt{LoLLiPoP} and \texttt{HiLLiPoP}), with respect to the more used likelihoods (PR4 \texttt{CamSpec} and PR3), is considered, the lensing anomaly is reduced, and the preference for $A_L>1$ and $\Omega_k<0$ is less significant. From the CMB+BAO+SNIa dataset, in the context of the parameterization $w_0w_a$CDM, we find a preference for a time-evoling dark energy over the rigid cosmological constant which is consistent with the most recent results from DESI collaboration.
\end{abstract}







\end{frontmatter}


\section{\label{sec:intro} Introduction:\protect} 
The combination of different cosmological probes during the last decades has led to the rise of a standard model of cosmology, the $\Lambda$CDM. This model can describe simultaneously the early universe from cosmic microwave background radiation \cite{WMAP:2012nax,Planck:2018vyg,ACT:2025fju}, the observed expansion rate evolution with redshift from type Ia Supernova (SN) probes \cite{SDSS:2014iwm,Brout:2022vxf,Rubin:2023ovl,DES:2024jxu}, baryonic acoustic oscillations (BAO) measurements \cite{2dFGRS:2001csf,SDSS:2005xqv,Gaztanaga:2008xz,Blake:2011wn,Beutler:2011hx,eBOSS:2020yzd,DESI:2024mwx} and lately the study of large-scale structure with the 3x2pt probes via weak lensing \cite{DES:2017myr,KiDS:2020suj,DES:2021wwk}.
Despite its success, there are some observational problems that the cosmological standard model is facing up. In particular, some direct measurements of the local expansion rate and indirect measurements from early universe probes show a tension between measurements of the Hubble-Lemaitre constant \cite{Cuesta:2014asa,Bernal:2016gxb,Verde:2019ivm,DiValentino:2021izs}. Other yet not resolved problems are: the radio cosmic dipole problem \cite{Oayda:2024hnu}, the $\sigma_8$ tension \cite{Chen:2022jzq} or the tension between 2D and 3D BAO measurements \cite{Anselmi:2018vjz,Favale:2024sdq}. More recently, the first sets of cosmological results with BAO measurements by the Dark Energy Spectroscopic Instrument Collaboration (DESI) \cite{DESI:2024mwx,DESI:2025zgx} and the final release of BAO and SN cosmological results from the Dark Energy Survey (DES) Collaboration \cite{DES:2024pwq,DES:2024jxu} have shown some preference for an evolving dark energy universe \cite{DES:2025bxy}. For earlier works claiming evidence in favor of a dynamical dark energy component see \cite{Sola:2015wwa,Sola:2016jky,SolaPeracaula:2016qlq,SolaPeracaula:2017esw,Zhao:2017cud,Park:2024vrw,Park:2024pew,Gomez-Valent:2024tdb}. For general reviews on the cosmological tensions and $\Lambda$CDM anomalies, see \cite{Perivolaropoulos:2021jda,Abdalla:2022yfr,Peebles:2022akh,CosmoVerse:2025txj}. Because of these anomalies and tensions, in the last years there has been also a growth on model-independent test studies of datasets for particular cosmological models \cite{LHuillier:2016mtc,Zhao:2017cud,Shafieloo:2018gin,LHuillier:2024rmp,Favale:2023lnp}. %
On the observational side, more observations are needed. There is a strong effort regarding the measurement of $H_0$ using the distance ladder, mainly by two projects, SH0ES program \cite{Riess:2021jrx}, that uses cepheids for calibration, and CCHP program, that uses three different calibrators (Tip of the red giant branch: TRGB, J-region asymptotic giant branch: JAGB, cepheids) \cite{Freedman:2024eph}. Alternative observational methods involve the use of gravitational wave binaries or strong lensing as they have different systematics \cite{LIGOScientific:2017adf,H0LiCOW:2019pvv,TDCOSMO:2025dmr}. Besides the Hubble constant, some high redshift measurements of the expansion rate have shown some hints of potential deviation from a cosmological constant $\Lambda$ such as with quasars \cite{Risaliti:2015zla,Lusso:2020pdb}, type Ia SN \cite{DES:2024jxu} and BAO \cite{DESI:2024mwx}. Of special importance are single rung measurements such as standard speed-guns \cite{Hodgson:2020tuw,Hodgson:2023mqs} because they allow us to reduce the level of dependence from calibrators.
Regarding the solution from possible theoretical solutions, some simple phenomenological extensions to $\Lambda$CDM have been proposed as an explanation for these tensions and anomalies. For example, the lensing anomaly induces a change on the amount of matter in the universe (through a change on the small scales) \cite{Calabrese:2008rt}. 
The latest official Planck Collaboration likelihood is called PR3 \cite{Planck:2018tab} and it has been considered the standard Planck likelihood.
Recently, a unified framework (NPIPE) has been applied to both the low frequency maps and high frequency maps of Planck mission \cite{Planck:2020olo} that also includes extra data collected by satellite repointings that were not included in the previous releases. The NPIPE maps have been used to build two different approaches to the Planck likelihoods for cosmological parameter estimation, the PR4-\texttt{CamSpec} \cite{Rosenberg:2022sdy} and the combined PR4-\texttt{LoLLiPoP} and PR4-\texttt{HiLLiPoP}  likelihoods \cite{Tristram:2023haj}. Some studies have recently used the PR4-\texttt{LoLLiPoP} and PR4-\texttt{HiLLiPoP} in combination with the newest DESI data to update cosmological parameter constraints on for a range of cosmological models \cite{RoyChoudhury:2024wri,RoyChoudhury:2025dhe}. The PR4-\texttt{CamSpec} likelihood for the high-$\ell$ multipoles of CMB power spectra has been extensively combined with other cosmological probes and for example the DESI Collaboration uses it as default CMB likelihood for the dark energy studies.  

%
However, the combination of the latest Planck likelihood (PR4-\texttt{LoLLiPoP} and PR4-\texttt{HiLLiPoP}) with BAO and SN measurements to study the tensions between the simplest $\Lambda$CDM extensions (e.g. Universe spatial curvature as a free parameter) has not been that extensively studied as deeply in the past as other dataset combinations.

%
In this paper, we use data from PR4 Planck \texttt{LoLLiPoP} and \texttt{HiLLiPoP} likelihood, DESI 2024 BAO data and SNIa from the Pantheon+ sample in order to get constraints in 5 models of cosmology to study the tensions in the parameter constraints. Throughout the paper we refer to PR4 Planck \texttt{LoLLiPoP} and \texttt{HiLLiPoP} likelihood as PR4.

\section{Data}\label{sec:data}
To constrain the cosmological parameters that characterize each of the models under study we utilize CMB anisotropy and weak lensing data, BAO and SNIa data. Some details of the different datasets, together with the corresponding references are provided in the following.
\newline
\newline 
{\bf PR4}: We consider the PR4 updated versions, with respect to the PR3 (which referes to the Planck 2018 TTTEEE data \cite{Planck:2018vyg}), of the likelihoods that contain the Planck CMB temperature anisotropy and polarization data as well as its corresponding cross-spectra \cite{Tristram:2023haj}. We refer the readers to that reference, where all the technical details involved in getting the upgraded versions of the \texttt{HiLLiPoP} (hlp) and \texttt{LoLLiPoP} (lol) are provided. In the CMB data combination considered only the low multipole temperature (lowT) likelihood comes from PR3. Following author's notation, the full TTTEEE likelihood is composed by: lowT ($2\leq \ell \leq 30$) + lolE ($2\leq \ell \leq 30$) + hlpTTTEEE ($30\leq \ell \leq 2500$ for TT and for TE and EE $30\leq \ell \leq 2000$). 
\newline
\newline
{\bf (PR4) lensing}: The lensing potential power spectrum obtained from the analysis of the Planck CMB PR4 maps \cite{Carron:2022eyg}.
\newline
\newline
{\bf BAO}: We employ a total of 12 BAO data points from both anisotropic and isotropic analyses from DESI 2024 \cite{DESI:2024mwx} spanning the redshift range $0.295 \leq z \leq 2.3330$. The 10 anisotropic data points are embodied in the $D_{\text{M}}(z_{\text{eff}})/r_d$ and $D_{\text{H}}(z_{\text{eff}})/r_d$ estimators whereas the 2 isotropic data points are given in the form of the $D_{\text{V}}(z_\text{eff})/r_d$ observable. The values for the different points together with the correlation coefficients for the anisotropic results are provided in Table 1 of \cite{DESI:2024mwx}. In this work we stick to the use of the 3D BAO data. We recommend \cite{Favale:2024sdq,Gomez-Valent:2024tdb,Gomez-Valent:2024ejh} for discussions about the tension between the results obtained with 3D BAO and 2D BAO data. %
\newline
\newline
{\bf SNIa}: The Pantheon+ compilation \cite{Brout:2022vxf}, contains a total of 1701 measurements of the apparent magnitude of supernova type Ia as a function of the redshift. In order to avoid dependencies of the considered model to deal with the peculiar velocities we remove from the likelihoods all the data points with redshift $z<0.01$. The remaining 1590 measurements probe the redshift range $0.01016\leq z \leq 2.26137$. In our analyses the absolute magnitude of supernova, usually denoted by $M$, is treated as a nuisance parameter. 

\section{\label{sec:methods} Methodology}
The $\Lambda$CDM model can be described by six cosmological parameters $\Omega_b h^2$, $\Omega_c h^2$, $H_0$, $\tau$, $A_s$ and $n_s$. The $\Omega_b$ and $\Omega_c$ represent the present values of the baryon and cold dark matter density parameters, $H_0$ is the current value of the Hubble function, $\tau$ is the reionization optical depth and the last two parameters are related with the scalar type primordial power spectrum
\begin{equation}
P_\delta(k) = A_s\left(\frac{k}{k_0}\right)^{n_s}    
\end{equation}
with $A_s$ the amplitude of the power spectrum, $n_s$ the spectral index and $k_0=0.05\text{Mpc}^{-1}$ the pivot scale. As our goal is to check the tensions and anomalies between different datasets and re-analysis, we study the standard model of cosmology, the so-called $\Lambda$CDM model together with some commonly studied extensions.
The background evolution of all the models under study can be generally described by the following expression of the Hubble function
\begin{equation}
\begin{split}
H^2(a) = H^2_0[\Omega_\gamma{a^{-4}} &+ (\Omega_b + \Omega_c){a^{-3}} \\
       &+ \Omega_k{a^{-2}} + \Omega_\nu(a)+ \Omega_{\text{DE}}(a)].
\end{split}
\end{equation}
where $\Omega_\gamma$ represents the present value of the photon density parameter, $\Omega_k$ is the curvature parameter, $\Omega_\nu(a)$ encodes the contribution from massive and massless neutrinos and $\Omega_{\text{DE}}(a)$ contains the contribution of the dark energy fluid. 
\newline
\newline
{\bf $\Lambda$CDM+$\Omega_k$}: In this model, the spatial hypersurfaces are not assumed to be flat. This implies changes in the calculation of cosmological distances, as well as modifications in the shape of the power spectrum of primordial fluctuations where we assume the same parametrization as the one used by the Planck Collaboration \cite{Planck:2018vyg,deCruzPerez:2022hfr}:
\begin{equation}
P_\delta(q)\propto\frac{(q^2-4K)^2}{q(q^2-K)^2}A_s\left(\frac{k}{k_0}\right)^{n_s-1}   
\end{equation}
where $q^2=k^2+K^2$ is the wave-number for the non-flat $\Lambda$CDM model and $K=-(H_0^2/c^2)\Omega_k$ accounts for the curvature. See \cite{Sanz-Wuhl:2024uvi} for the role of the assumed value of $\Omega_k$ in the fiducial cosmology.
\newline
\newline
{\bf $\Lambda$CDM+$A_L$}: The CMB photons go through weak lensing scattering during their travel through the large-scale structures of the universe. The CMB lensing is described \cite{Hu:2000ee,Lewis:2006fu} by the power spectrum of the lensing potential $C_\ell^{\psi\psi}$ and the impact on the TT power spectrum of CMB is given by a convolution with the unlensed TT power spectrum $C_\ell$. In order to check the consistency of the CMB lensing models, within the context of the $\Lambda$CDM, a test parameter $A_L$ was introduced \cite{Calabrese:2008rt} in the lensing power spectrum  as $C_\ell^{\psi\psi}\rightarrow A_LC_\ell^{\psi\psi}$ so the expression for the lensed CMB TT, $\tilde{C}_\ell$, power spectrum is given by:
%
\begin{equation}
    \tilde{C}_\ell \approx C_\ell + A_L \int \frac{d^2 \boldsymbol{\ell'}}{(2\pi)^2} [\boldsymbol{\ell'}\cdot(\boldsymbol{\ell}-\boldsymbol{\ell'})]^2 C_{|\boldsymbol{\ell} - \boldsymbol{\ell}'|}^{\psi\psi} C_{\ell'}.
\end{equation}
where $A_L=1$ if the modeling of $C_{|\boldsymbol{\ell} - \boldsymbol{\ell}'|}^{\psi\psi}$ within general relativity is correct.
Allowing the phenomenological parameter $A_L$ to vary in our analysis will allow us to check whether the theoretical prediction for the amount of weak lensing in the CMB spectra is correct or not. 
\newline
\newline
{\bf $w_0$CDM}: This is a simple zero-order parametrization 
that allows to track a possible time-evolution of the dark energy component. The parameterization is characterized by a constant dark energy equation of state parameter $w_0$ whose value may in principle deviate from the cosmological constant behavior $w_0\neq -1$. This can mimic the quintessence ($w_0>-1$) or phantom ($w_0<-1$) behavior of a scalar field model, provided that the corresponding equation of state parameter is approximately constant. The cosmological constant contribution is replaced by $\Omega_{\text{DE}}(a) = \Omega_\Lambda a^{-3(1+w_0)}$.  
\newline
\newline
{\bf $w_0{w_a}$CDM}: This model is the next order parametrization, usually named as Chevallier-Polarski-Linder (CPL) \cite{Chevallier:2000qy,Linder:2002et}, with respect to $w_0$CDM and accounts for the possibility of a scale factor dependence of the dark energy equation of state parameter $w(a) = w_0 + w_a(1-a)$, which has a well-defined asymptotic limit $w(a\rightarrow0)\simeq w_0+w_a$ in the early universe. In this case the dark energy density reads $\Omega_{\text{DE}}(a)=\Omega_\Lambda a^{-3(1+w_0+w_a)}e^{-3w_a(1-a)}$. Check \cite{Li:2019yem,Park:2025azv,DESI:2025fii,Gonzalez-Fuentes:2025lei,Colgain:2025nzf} for alternative dark energy parameterizations. 
\newline
\newline
The performance of the different cosmological models and parameterizations, when it comes to fitting the observation data, is studied through the computation of the joint log-likelihood $\log{\mathcal{L}}$. Assuming the relation $\chi^2\propto -2\log{\mathcal{L}}$, the joint $\chi^2$-function, whose individual components are 
\begin{equation}
\chi^2_{\text{tot}} = \chi^2_{\text{CMB}}+\chi^2_{\text{lensing}}+\chi^2_{\text{BAO}}+\chi^2_{\text{SNIa}}    
\end{equation}
where here, for the sake of simplicity, CMB stands for the TTTEEE likelihood described in the previous section.
\newline
\newline
The cosmological background and perturbation equations are solved by making use of the Einstein-Boltzmann system solver \texttt{CLASS} \cite{Lesgourgues:2011re,Blas:2011rf} whereas the exploration of the parameter space is carried out through the use of a Markov chain Monte Carlo analysis being the corresponding algorithm implemented in \texttt{Cobaya} \cite{Torrado:2020dgo}. As a criterion to ensure that convergence has been reached we use the Gelman-Rubin convergence statistics \cite{R1:1997,R2:1992} setting the condition to stop the Monte Carlo sampling when $R-1<0.01$. Once the converged chains are obtained we employ the code \texttt{GetDist} \cite{Lewis:2019xzd} to get the corresponding posterior distributions. 
In our analyses we consider conservative flat priors, in particular, for the six primary parameters common to all the models: $0.005\leq\Omega_b h^2\leq 0.1$, $0.001\leq\Omega_c h^2 \leq 0.99$, $20\leq H_0[\text{km/s/Mpc}]\leq 100$, $0.01\leq \tau\leq 0.8$, $0.8\leq n_s\leq 1.2$ and $1.61 \leq \ln(10^{10}A_s)\leq 3.91$. As for the non-standard parameters, while in the case of the $\Lambda$CDM+$\Omega_k$ model, the curvature parameter varies within the range $-0.5\leq \Omega_k\leq 0.5$, for the $\Lambda$CDM+$A_L$ the lensing parameter is allowed to vary within $0\leq A_L \leq 10$. In regard to the dynamical dark energy parameterizations, for $w_0$CDM  $-3.0 \leq \omega_0 < 0.2$ and for $\omega_0\omega_a$CDM $-3.0\leq \omega_0 \leq 0.2$ and $-3\leq \omega_a \leq 2$. In the tables, in addition to the values of the primary cosmological parameters, we also provide the values of three derived parameters, namely: the total non-relativistic matter density parameter $\Omega_m$ which contains the contribution of baryons, cold dark matter and massive neutrinos, $\sigma_8$, the root mean square of matter perturbations computed at the scale $R_8 = 8/h$Mpc \footnote{See \cite{Sanchez:2020vvb,Forconi:2025cwp} for discussions in favor of using $\sigma_{12}$ instead of $\sigma_8$.} and $r_d$, the radius of the sound horizon evaluated at the drag epoch $z_d$. Finally, we fix the current value of the CMB temperature to $T_0 = 2.7255$ K \cite{Fixsen:2009ug} and we consider three different neutrino species, being two of them massless and one massive with $m_\nu=0.06$ eV. Since the BAO+SNIa dataset alone is not capable of constraining the parameters $\tau$ and $n_s$, in the corresponding analyses, we fix their values to the ones obtained using the PR4 data. As mentioned before, the parameter $A_L$ is introduced to rescale the lensing power spectrum and consequently it plays a very minor role (only through correlations with other parameters) in the late time universe. Therefore in Table \ref{tab:results_LCDM_AL} we do not provide the results when only BAO+SNIa data are analyzed since the results are basically the same as the ones obtained for the $\Lambda$CDM model.
\newline
\newline
In order to compare the performance of the different models and parameterizations under study when it comes to fitting the cosmological data somehow we need to penalize the presence of extra parameters. This can be accomplished by making use of the deviance information criterion (DIC) \cite{DIC,Kunz:2006mc,Liddle:2007fy}, which value can be computed with 
\begin{equation}
\text{DIC} = \chi^2(\hat{\theta}) + 2p_D.     
\end{equation}
In the above expression $p_D = \overline{\chi^2} -\chi^2(\hat{\theta})$ represents the effective number of parameters (the number can be obtained from the values provided in corresponding tables), with $\overline{\chi^2}$ we denote the average value of the $\chi^2$-function and $\chi^2(\hat{\theta})$ is the $\chi^2$-function evaluated in the best-fit values of the fitting parameters. The best-fit values are obtained with the \texttt{BOBYQA} minimizer \cite{Cartis:2018xum,Cartis:2018jxl}. We define the difference between the $\chi^2_{\text{min}}$ of a given extended model $X$ and the $\Lambda$CDM one as $\Delta\chi^2_{\text{min}}=\chi^2_{\text{min},\Lambda\text{CDM}}-\chi^2_{\text{min},X}$. We compute differences in the DIC value with respect to the $\Lambda$CDM model as 
\begin{equation}
\Delta\text{DIC} = \text{DIC}_{\Lambda\text{CDM}} - \text{DIC}_{\text{X}}    
\end{equation}
where the X represents all the non-standard models and parameterizations considered in this work. If we get $\Delta \text{DIC}>0$ the non-standard model is favored over the $\Lambda$CDM whereas $\Delta \text{DIC} <0$ points out to a preference of the cosmological data for the standard model of cosmology. According to the usual standards, values $0\leq \Delta\text{DIC} <2$ would indicate {\it weak} evidence in favor of the non-standard model, if $2\leq \Delta\text{DIC} <6$ there would be {\it positive} evidence whereas for values in the range $6\leq \Delta\text{DIC} <10$ point out to {\it strong} preference for the cosmological extended model. Finally if we find values $\Delta\text{DIC}>10$ we are allowed to claim {\it very strong} evidence in favor of the non-standard model. 

To check how consistent are, two sets of cosmological parameter constraints obtained with two different dataset, within a given model we use another statistical estimator based on the \text{DIC} values \cite{Joudaki:2016mvz,deCruzPerez:2022hfr}. 
\begin{equation}
 \mathcal{I}(D_1,D_2)\equiv{\text{exp}{\left(-\frac{\mathcal{G}(D_1,D_2)}{2}\right)}}   
\end{equation} 
where $D_1$ and $D_2$ are the two datasets under comparison and we define
\begin{equation}
    \mathcal{G}=\text{DIC}(D_1\cup D_2)-\text{DIC}(D_1)-\text{DIC}(D_2).
\end{equation}
This estimator is built in such a way that values $\text{log}_{10}\mathcal{I}>0$ indicate consistency between the results obtained with the two datasets whereas $\text{log}_{10}\mathcal{I}<0$ points to an inconsistency. According to Jeffrey's scale, we claim that the level of concordance/discordance is {\it inconclusive} if $|\text{log}_{10}\mathcal{I}|<0.5$, {\it substantial} if $|\text{log}_{10}\mathcal{I}|>0.5$, {\it strong} when $|\text{log}_{10}\mathcal{I}|>1$ and {\it decisive} for $|\text{log}_{10}\mathcal{I}|>2$.

\section{\label{sec:results} Results}

\begin{figure}
\centering
\includegraphics[width=0.49\textwidth]{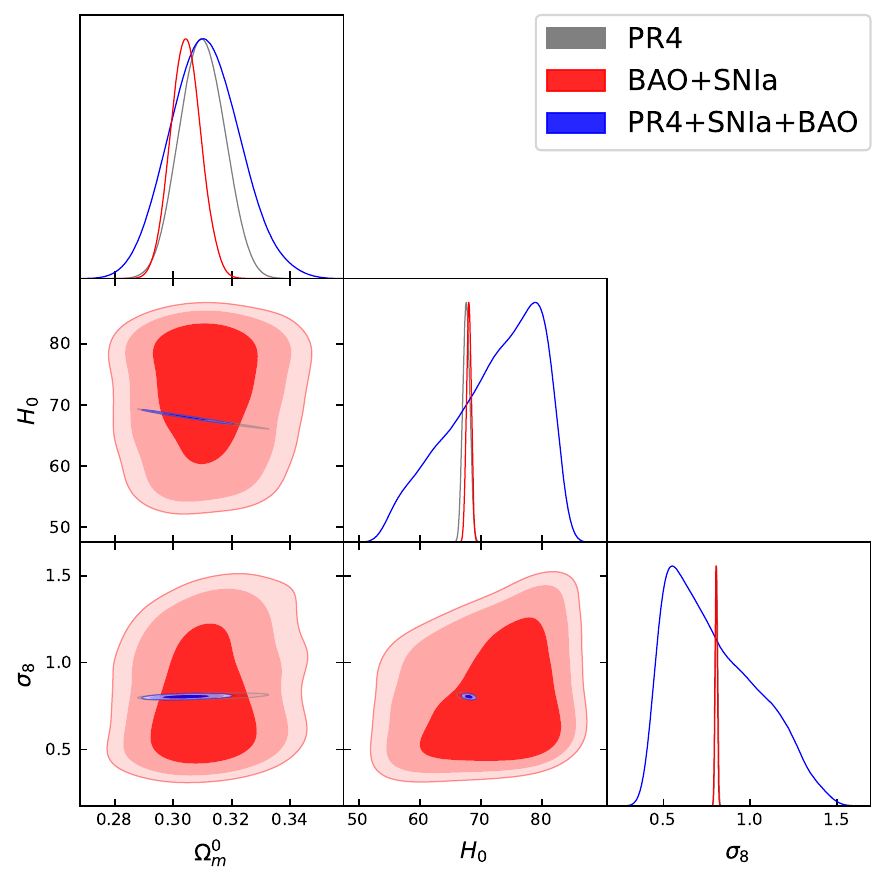}
\caption{{\bf $\Lambda$CDM}: Cosmological parameter constraints for the $\Lambda$CDM model obtained with the BAO+SNIa, PR4 and PR4+BAO+SNIa datasets. The parameter $H_0$ is expressed in km/s/Mpc units.}
\label{fig:LCDM}
\end{figure}

\begin{figure}
\centering
\includegraphics[width=0.49\textwidth]{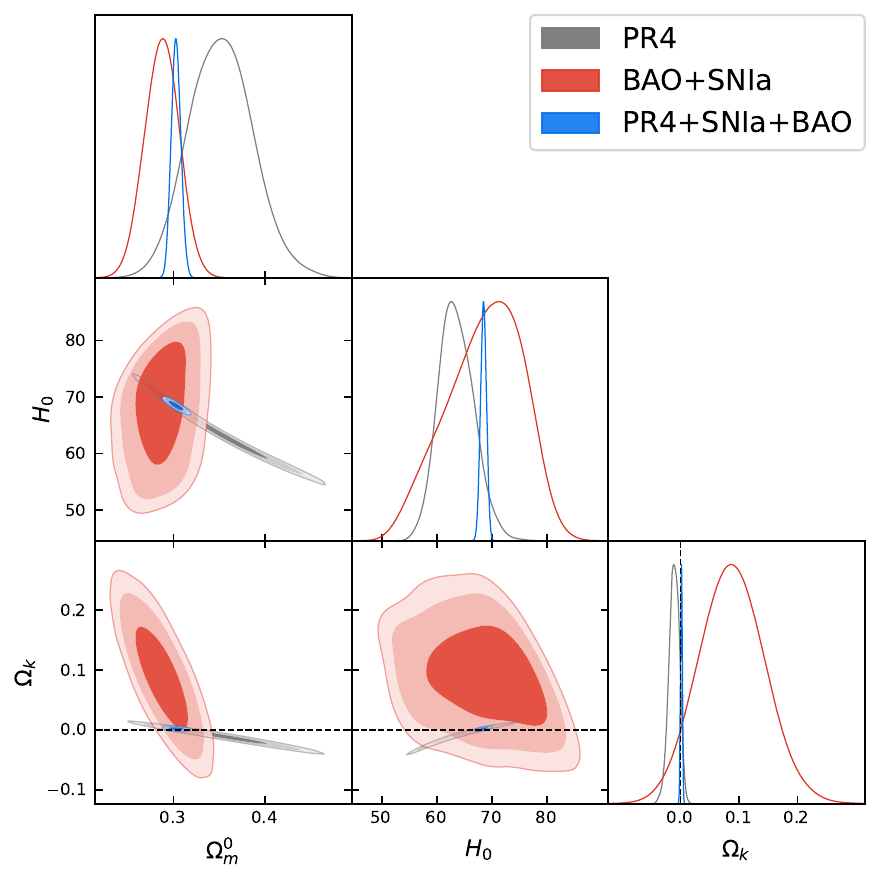}
\caption{{\bf $\Lambda$CDM+$\Omega_k$}: Cosmological parameter constraints for the $\Lambda$CDM+$\Omega_k$ model obtained with the BAO+SNIa, PR4 and PR4+BAO+SNIa datasets. The parameter $H_0$ is expressed in km/s/Mpc units. We include the dashed lines to highlight the flat Universe case ($\Omega_k=0$).}
\label{fig:OkCDM}
\end{figure}

\begin{figure}
\centering
\includegraphics[width=0.49\textwidth]{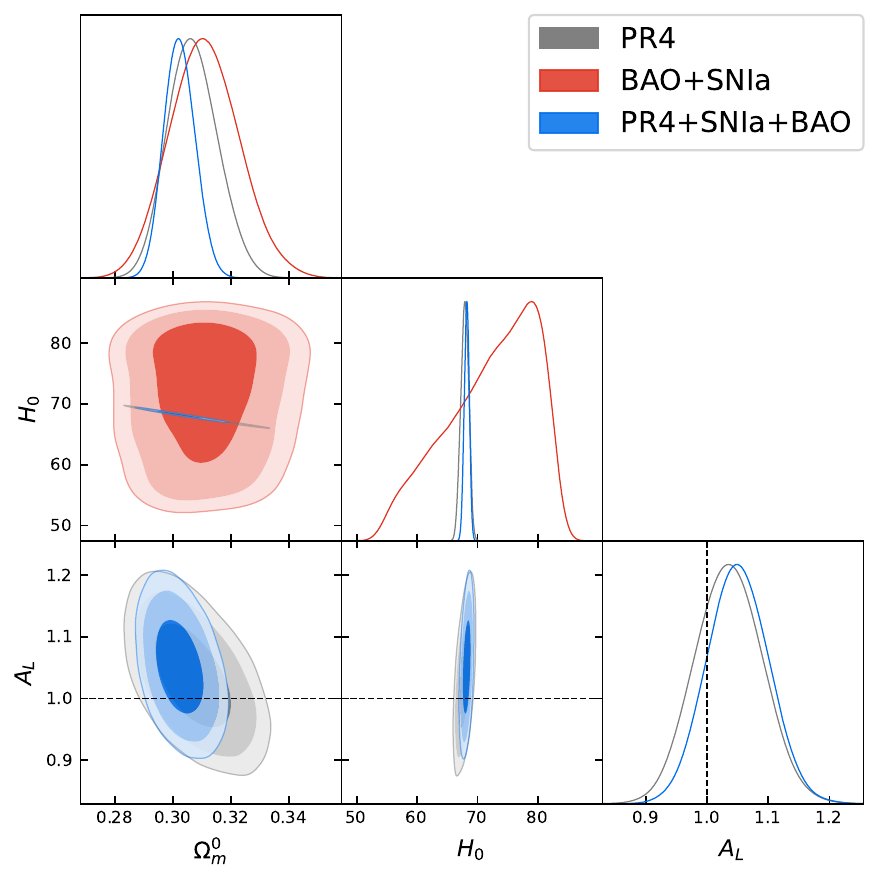}
\caption{{\bf $\Lambda$CDM+$A_L$}: Cosmological parameter constraints for the $\Lambda$CDM+$A_L$ model obtained with the BAO+SNIa, PR4 and PR4+BAO+SNIa datasets. The parameter $H_0$ is expressed in km/s/Mpc units. The dashed lines represent the $A_L=1$ case.}
\label{fig:LCDM_AL}
\end{figure}

\begin{figure}
\centering
\includegraphics[width=0.49\textwidth]{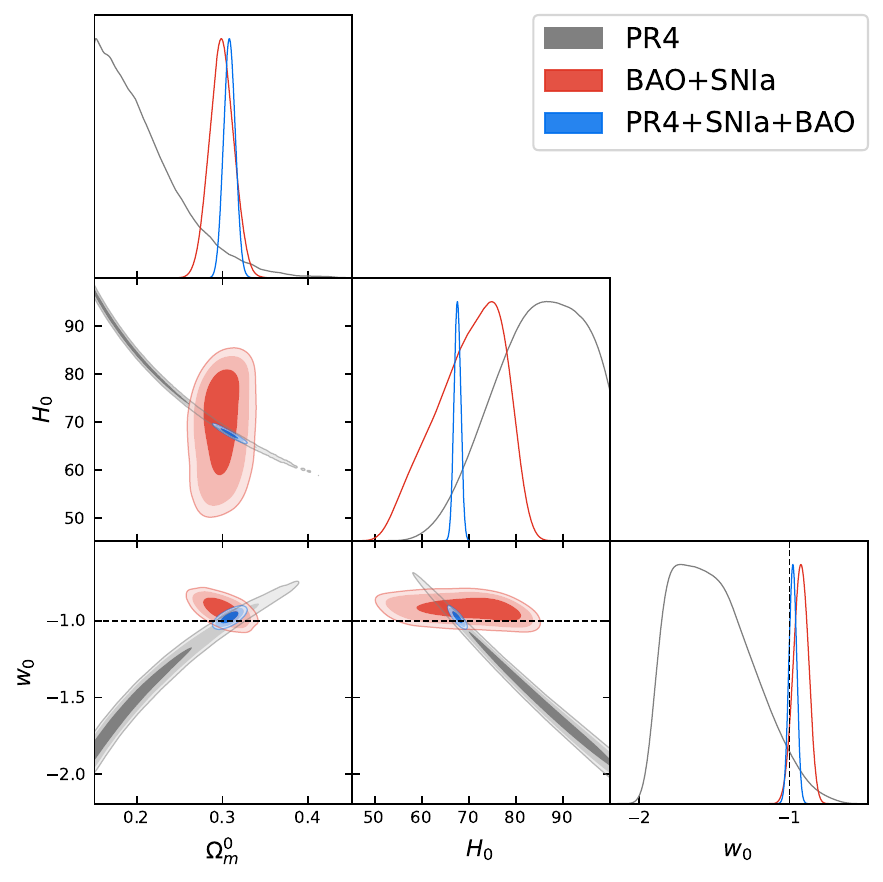}
\caption{{\bf $w_0$CDM}: Cosmological parameter constraints for the $w_0$CDM model obtained with the BAO+SNIa, PR4 and PR4+BAO+SNIa datasets. The parameter $H_0$ is expressed in km/s/Mpc units. The dashed lines highlight the $\Lambda$CDM case.}
\label{fig:XCDM}
\end{figure}

\begin{figure}
\centering
\includegraphics[width=0.49\textwidth]{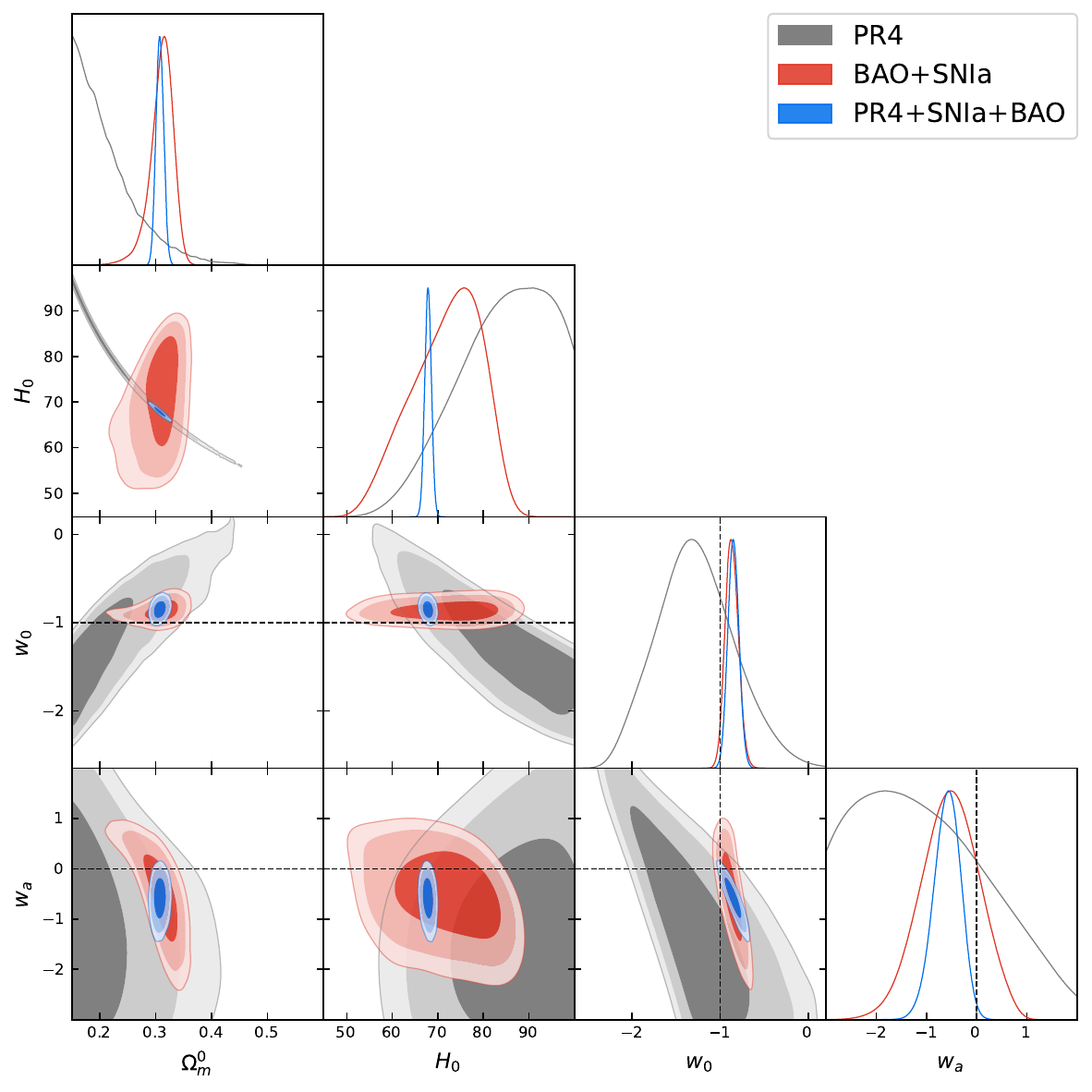}
\caption{{\bf $w_0w_a$CDM}: Cosmological constraints for the $w_0w_a$CDM model obtained with the BAO+SNIa, PR4 and PR4+BAO+SNIa datasets. The parameter $H_0$ is expressed in km/s/Mpc units. The dashed lines in the plot are a reference to the $\Lambda$CDM case.}
\label{fig:CPL}
\end{figure}

The cosmological parameter constraints obtained with the datasets:\\ BAO+SNIa, PR4, PR4+lensing, PR4+BAO+SNIa and PR4+lensing+BAO\\+SNIa for the models $\Lambda$CDM, $\Lambda$CDM+$\Omega_k$, $\Lambda$CDM+$A_L$, $w_0$CDM and the $w_0w_a$CDM are provided in Tables \ref{tab:results_LCDM},\ref{tab:results_LCDM_Omega_k},\ref{tab:results_LCDM_AL},\ref{tab:results_w0CDM} and \ref{tab:results_w0waCDM} respectively. The corresponding two-dimensional contour plots can be seen in Figures \ref{fig:LCDM}, \ref{fig:LCDM_AL}, \ref{fig:OkCDM}, \ref{fig:XCDM} and \ref{fig:CPL}. 
The replacement of the PR3 CMB likelihood by the recent PR4 one brings significant changes in the results. The PR3 results displayed in this section are extracted from \cite{Planck:2018tab}. With the PR3 likelihood, when the $\Lambda$CDM+$\Omega_k$ is analyzed, a value of $\Omega_k=-0.044^{+0.018}_{-0.015}$ is obtained and the difference in the value of $\chi^2_{\text{min}}$ with respect to the $\Lambda$CDM model is $\Delta\chi^2_{\text{min}}=+11.36$. On the other hand, when we employ the PR4 likelihood to test the models, we obtain for the curvature model $\Omega_k = -0.0107^{+0.0096}_{-0.0080}$ and $\Delta\chi^2_{\text{min}}=+3.80$. While both likelihoods still show preference for a closed universe, the significance of the results is different, being the first value $2.44\sigma$ away from $\Omega_k=0$ whereas the second result favors non-flat spatial hypersurfaces by $1.11\sigma$. For model-independent measurements of the curvature parameter $\Omega_k$ see \cite{Favale:2023lnp} or alternative models \cite{Deng:2024uuz} compatible with this result. In regard to the $\Lambda$CDM+$A_L$ model, results along the same lines are obtained. When the PR3 data are analyzed, the favored value for the lensing parameter is $A_L = 1.180\pm 0.065$ which deviates from $A_L=1$ by $2.77\sigma$ being the difference in the minimum value of the $\chi^2$-function with respect to the standard model $\Delta\chi^2_{\text{min}}=+9.66$, whereas when the PR4 data are considered the results are $A_L=1.036\pm 0.055$ ($0.65\sigma$) and $\Delta\chi^2_{\text{min}}=+2.53$. In light of these results one may claim that the so-called lensing anomaly is less significant once the PR3 likelihood is replaced in the analyses by the PR4 one. It is also interesting to see how the results change when the lensing data are added to the mix. In the case of the $\Lambda$CDM+$\Omega_k$ when the PR3+lensing data are considered (being the lensing likelihood the one from PR3), the result obtained are $\Omega_k =-0.0106\pm 0.065$ and $\Delta\chi^2_{\text{min}}=+3.22$, representing a considerable reduction, when compared to the PR3 alone case, not only in the mild evidence in favor of a closed universe ($1.63\sigma$) but also in the better performance with respect to the standard model when it comes to fitting the data. As for the PR4+lensing analysis we get $\Omega_k=-0.0073^{+0.0063}_{-0.0052}$ ($1.16\sigma$) and $\Delta\chi^2_{\text{min}}=+1.99$, which indicates more stability in the results when we move from PR4 to PR4+lensing data. A similar trend is observed in the $\Lambda$CDM+$A_L$ but with some minor differences that are worth to be mentioned. While in the PR3+lensing analysis the results are $A_L = 1.071^{+0.038}_{-0.042}$ ($1.69\sigma$) and $\Delta\chi^2_{\text{min}}=+3.43$ in the PR4+lensing case we have $A_L = 1.041\pm 0.038$ ($1.10\sigma$) and $\Delta\chi^2_{\text{min}}=+3.08$. Therefore, when PR3 data are considered there exists a clear decrease in the evidence in favor of $A_L>1$ when the lensing data are included. However, when PR4+lensing data are jointly analyzed, the option $A_L>1$ is more favored in comparison with the case with only PR4 data. In regard to the dynamical dark energy parameterizations $w_0$CDM and $w_0 w_a$CDM, even if the lensing data contains information about the late-time universe, either PR3 (PR4) or PR3+lensing (PR4+lensing), are not able to put tight constraints over the equation of state parameters and consequently no significant differences are appreciated when we move from PR3 to PR4. 
\newline 
\newline 
For all the models under study we do not appreciate significant changes in the cosmological parameter constraints when we compare PR4+BAO+SNIa results and PR4+lensing+BAO+SNIa ones, therefore, we focus on the discussion of the results obtained with the second datasets since it is the most complete one. 
When the curvature paramater $\Omega_k$ is allowed to vary, we find for the $\Lambda$CDM+$\Omega_k$ model, $\Omega_k=0.0018\pm 0.0015$, indicating now a $1.2\sigma$ preference for an open universe. This somehow is in contradiction with the results obtained with PR4 for the same model, where a preference for a closed universe was found. This is reflected in the value $\text{log}_{10}\mathcal{I}=-0.99$, obtained when we compare PR4 vs. BAO+SNIa results,  which is on the verge of indicating a strong discordance between PR4 and BAO+SNIa results. This can also be seen by taking a look at Figure \ref{fig:OkCDM} where non-overlaping contour plots at $\sim 2\sigma$ can be observed. In the comparison with the $\Lambda$CDM we observe a {\it positive} evidence in favor of the standard model. For the $\Lambda$CDM+$A_L$ model, we get $A_L = 1.050\pm 0.035$ which represents a deviation of $1.43\sigma$ with respect to the $A_L=1$ expected value. The presence of the varying $A_L$ parameter allows the model to slightly improve the performance of the $\Lambda$CDM ($\Delta\chi^2_{\text{min}}=+2.10$), however according to the DIC, $\Delta\text{DIC}=+0.92$ the model is only {\it weakly} preferred over the standard model. As for the dynamical dark energy parameterizations we have for the $w_0$CDM, a value of the equation of state parameter $w_0 =-0.984\pm 0.026$ which shows a $0.62\sigma$ deviation from the value $w_0=-1$ associated to the cosmological constant. Which such small deviation with respect to the $w=-1$ this parameterization does not improve the $\Lambda$CDM fit. As expected the minimum value of the $\chi^2$ function is below the one found for the $\Lambda$CDM, in particular $\Delta \chi^2_{\min}=+0.76$, however, when we penalize the presence of extra parameters $\Delta \text{DIC}=-1.58$ it turns out that it is the standard model the one that is {\it weakly} preferred over the $w_0$CDM parameterization. According to the contour plots displayed in Figure \ref{fig:XCDM}, there is tension between the cosmological parameter constraints obtained with PR4 and BAO+SNIa. This is also reflected in the value of the equation of state parameter, that points to a phantom behavior $w_0=-1.51^{+0.18}_{-0.35}$ when PR4 data are analyzed and quintessence $w_0=-0.978\pm 0.026$ when PR4+BAO+SNIa are jointly analyzed. However, this tension is not fully captured in the statistical estimator $\text{log}_{10}\mathcal{I}=-0.15$, which is not conclusive. On the other hand, the $w_0w_a$CDM parameterization is favored when compared to the $\Lambda$CDM. For the equation of state parameters we obtain $w_0 = -0.843\pm 0.062$ and $w_a= -0.64^{+0.29}_{-0.19}$. A more suitable parameter combination to check deviations from the $\Lambda$CDM ($w=-1$) is $w_0+w_a$, which is the value of the equation of state in the limit of small values of $a$. If there is no deviation with respect to standard model then we should find $w_0+w_a=-1$, but we get $w_0+w_a=-1.48^{+0.23}_{-0.19}$ which is $2.1\sigma$ away from $w=-1$. We also obtain $\chi^2_{\min}=+5.31$ and $\Delta \text{DIC}=+2.99$ which indicates that the $w_0 w_a$CDM parameterization is {\it positively} favored when compared to the $\Lambda$CDM model. As for the concordance between PR4 and BAO+SNIa, according to the estimator employed in this work, we find $\text{log}_{10}\mathcal{I}=1.08$ pointing to a good agreement between the results obtained with the two datasets. It is interesting to note that similar results were found in \cite{deCruzPerez:2024shj,Park:2024vrw,Park:2024pew}, where PR3 and non-DESI data are used, which indicates that the results are stable regardless of the dataset considered. When using the same datasets and parameter extensions, our results are agree with the results in \cite{RoyChoudhury:2024wri}.
\section{Conclusions}\label{sec:conclusions}
In this work we have tested some of the most common phenomenological extensions of the $\Lambda$CDM model in order to see whether they are capable of improving the performance of the standard model and we have reevaluated the status of the lensing anomaly by using PR4 CMB data and BAO DESI 2024 data. When PR3 data are replaced by PR4 data the evidence in favor of $\Omega_k<0$ and $A_L>1$ is clearly reduced. In particular we find $\Omega_k=-0.0107^{+0.0096}_{-0.0080}
$ and $A_L=1.036\pm 0.055$ indicating a deviation of $1.11\sigma$ and $0.72\sigma$ from $\Omega_k=0$ and $A_L=1$, respectively. Therefore, we may claim that when PR4 data are analyzed the lensing anomaly almost subsides. Remarkably, the inclusion of the PR4-lensing data and the combination of BAO+SNIa increases the preference for non-flat spatial hypersurfaces ($\Omega_k=0.0018\pm 0.0015$) and for $A_L\neq 1$ ($A_L=1.050\pm 0.035$). When PR4+lensing+BAO+SNIa data are analyzed, for the dynamical dark energy parameterization $w_0$CDM we find a slight shift ($0.61\sigma$) for a quintessence behavior $w_0=-0.984\pm 0.026$, however, according to the statistical criterion DIC, the model is {\it weakly} disfavored when compared to the $\Lambda$CDM. On the other hand for the $w_0 w_a$CDM parameterization, when the same dataset is considered, we obtain $w_0+w_a=-1.48^{+0.23}_{-0.19}$, which is separated by $\sim 2\sigma$ from the value expected $w_0+w_a=-1$ in the standard model. In regard to the performance when it comes to fitting the data, we obtain $\Delta\text{DIC}=+2.99$ indicating that the model is {\it positively} preferred over the $\Lambda$CDM model. Although these results are obtained within the context of a mere parameterization, this could serve as a clue to the behavior that the equation of state parameter, of more complex physical models, should follow in order to improve the performance of the standard model of cosmology.  
\newline 
\newline 
%
%
\begin{table*}
\begin{adjustbox}{width=1\textwidth}
\small
\begin{tabular}{lccccc}

  Parameter                     &  BAO+SNIa                     & PR4                         &  PR4+lensing               &  PR4+BAO+SNIa        & PR4+lensing+BAO+SNIa     \\[+1mm]
 \hline \\[-1mm]
  $\Omega_b h^2$                & $0.0226^{+0.0120}_{-0.0050}$          & $0.02223 \pm 0.00014$       & $0.02221 \pm 0.00014$      &  $0.02229 \pm 0.00012$  &  $0.02229 \pm 0.00012$  \\[+1mm]
  $\Omega_c h^2$                & $0.137^{+0.030}_{-0.022}$  & $0.1188\pm 0.0013$         & $0.1192 \pm 0.0012$        &  $0.11795 \pm 0.00088$    &  $0.11817 \pm 0.00082$  \\[+1mm]
  $H_0$       & $72^{+10}_{-5}$  & $67.63 \pm 0.56$       & $67.49 \pm 0.52$      &  $68.04 \pm 0.39$  &  $67.95 \pm 0.37$  \\[+1mm]
  $\tau$                        & $0.0579$                     & $0.0579\pm 0.0063$         & $0.0588\pm 0.0062$        &  $0.0591\pm 0.0063$    &  $0.0608\pm 0.0061$    \\[+1mm]
  $\ln(10^{10} A_s)$                         & $2.76\pm 0.63$                     & $3.039 \pm 0.014$         & $3.045\pm 0.012$        &  $3.040 \pm 0.014$    &  $3.049 \pm 0.012$    \\[+1mm]
  $n_s$            & $0.9671$   & $0.9671 \pm 0.0042$           & $0.9664 \pm 0.0041$          &  $0.9694 \pm 0.0036$      &  $0.9689 \pm 0.0035$      \\[+1mm]
  \hline \\[-1mm]
  $\Omega_m$                           & $0.311\pm 0.012$           & $0.3099\pm 0.0077$     & $0.3119\pm 0.0071$    &  $0.3043\pm 0.0052$     &  $0.3056 \pm 0.049$     \\[+1mm]  
  $\sigma_8$                           & $0.78^{+0.16}_{-0.34}$           & $0.8061\pm 0.0066$     & $0.8093\pm 0.0051$    &  $0.8039\pm 0.0063$     &  $0.8081 \pm 0.0050$     \\[+1mm]    
  $r_d$                         & $140.6^{+6.6}_{-19}$       & $147.56\pm 0.27$ & $147.49\pm 0.25$ &  $147.73\pm 0.21$&  $147.67\pm  0.20$     \\[+1mm]   
  \hline\\[-1mm]
  $\chi^2_\text{min}$                         & $1418.28$       & $30551.10$ & $30559.38$ &  $31969.59$&  $31978.16$     \\[+1mm]  
  $\text{DIC}$                         & $1422.30$       & $30599.64$ & $30608.98$ &  $32021.87$&  $32031.64$     \\[+1mm]
  $p_D$                         & $2.01$       & $24.27$ & $24.80$ &  $26.14$&  $26.74$     \\[+1mm]
\end{tabular}
\\[+1mm]
\end{adjustbox}
\caption{{\bf $\Lambda$CDM}: Mean and 68\% confidence limits of the primary and derived parameters of the $\Lambda$CDM model obtained after analyzing the datasets: BAO+SNIa, PR4, PR4+lensing, PR4+BAO+SNIa and PR4+lensing+nonCMB. The $H_0$ has units of km/s/Mpc whereas $r_d$ has units of Mpc. We also include the values of $ \chi^2_{\text{min}}$,DIC and $p_D$.}
\label{tab:results_LCDM}
\end{table*}



\begin{table*}	
\begin{adjustbox}{width=1\textwidth}
\small
\begin{tabular}{lccccc}

  Parameter                     &  BAO+SNIa                     & PR4                         &  PR4+lensing               &  PR4+BAO+SNIa        & PR4+lensing+BAO+SNIa     \\[+1mm]
 \hline \\[-1mm]
  $\Omega_b h^2$                & $0.0266^{+0.012}_{-0.0049}$          & $0.02229\pm 0.00014$       & $0.02228\pm 0.00014$      &  $0.02221\pm 0.00013$  &  $0.02221 \pm 0.00013$  \\[+1mm]
  $\Omega_c h^2$                & $0.110^{+0.020}_{-0.028}$  & $0.1183\pm 0.0013$         & $0.1183\pm 0.0013$        &  $0.1190 \pm 0.0012$    &  $0.1192 \pm 0.0012$  \\[+1mm]
  $H_0$       & $69^{+8}_{-6}$  & $63.6^{+3.0}_{-3.5}$       & $64.7\pm 2.2$      &  $68.46 \pm 0.54$  &  $68.36 \pm 0.53$  \\[+1mm]
  $\tau$                        & $0.0564$                     & $0.0564\pm 0.0062$         & $0.0567\pm 0.0063$        &  $0.0579\pm 0.0063$    &  $0.0598\pm 0.0061$    \\[+1mm]
  $\ln(10^{10} A_s)$                         & $2.75\pm 0.62$                     & $3.035\pm 0.014$         & $3.035\pm 0.014$        &  $3.040\pm 0.014$    &  $3.049\pm 0.012$    \\[+1mm]
  $n_s$            & $0.9689$   & $0.9689\pm 0.0043$           & $0.9687\pm 0.0043$          &  $0.9667 \pm 0.0041$      &  $0.9664 \pm 0.0041$      \\[+1mm]
  $\Omega_k$            & $0.088\pm 0.056$   & $-0.0107^{+0.0096}_{-0.0080}$           & $-0.0073^{+0.0063}_{-0.0052}$          &  $0.0018 \pm 0.0015$      &  $0.0018\pm 0.0015$      \\[+1mm]
  \hline \\[-1mm]
  $\Omega_m$                           & $0.287\pm 0.019$           & $0.352\pm 0.035$     & $0.338^{+0.020}_{-0.023}$    &  $0.3028\pm 0.0052$     &  $0.3040 \pm 0.0051$     \\[+1mm]  
 $\sigma_8$                           & $0.62^{+0.14}_{-0.29}$           & $0.797\pm 0.010$     & $0.7990\pm 0.0096$    &  $0.8079\pm 0.0071$     &  $0.8120 \pm 0.0059$     \\[+1mm]  
  $r_d$                         & $147.2^{+8.7}_{-19}$       & $147.64\pm 0.27$ & $147.64\pm 0.27$ &  $147.53\pm 0.26$&  $147.49\pm 0.26$     \\[+1mm]  
 \hline\\[-1mm]
  $\chi^2_\text{min}$                         & $1415.20$       & $30547.30$ & $30557.39$ &  $31965.7$&  $31974.71$     \\[+1mm] 
  $\Delta\chi^2_\text{min}$                   & $+3.08$            & $+3.80$ & $+1.99$ &  $+3.89$&  $+3.45$     \\[+1mm]  
  $\text{DIC}$                         & $1421.24$       & $30601.10$ & $30608.97$ &  $32026.91$&  $32034.59$     \\[+1mm]  
  $p_D$                         & $3.02$       & $26.90$ & $25.79$ &  $30.61$&  $29.94$     \\[+1mm]
  $\Delta\text{DIC}$                         & $+1.06$       & $-1.46$ & $+0.01$ &  $-2.47$&  $-2.95$     \\[+1mm]  
\end{tabular}
\\[+1mm]
\end{adjustbox}
\caption{{\bf $\Lambda$CDM + $\Omega_k$}: Mean and 68\% confidence limits of the primary and derived parameters of the $\Lambda$CDM+$\Omega_k$ model obtained after analyzing the datasets: BAO+SNIa, PR4, PR4+lensing, PR4+BAO+SNIa and PR4+lensing+nonCMB. The $H_0$ has units of km/s/Mpc whereas $r_d$ has units of Mpc. We also include the values of $ \chi^2_{\text{min}}$ and DIC together with the differences with respect to the $\Lambda$CDM model, denoted by $\Delta\chi^2_{\text{min}}$ and $\Delta$DIC, respectively. We include $p_D$, the effective number of free parameters. }
\label{tab:results_LCDM_Omega_k}
\end{table*}


\begin{table*}
\begin{adjustbox}{width=1\textwidth}
\small
\begin{tabular}{lcccc}

  Parameter                                         & PR4                         &  PR4+lensing               &  PR4+BAO+SNIa        & PR4+lensing+BAO+SNIa     \\[+1mm]
 \hline \\[-1mm]
  $\Omega_b h^2$                         & $0.02228^{+0.00016}_{-0.00015}$       & $0.02229^{+0.00016}_{-0.00014}$      &  $0.02234^{+0.0013}_{-0.0012}$  &  $0.02235 \pm 0.00012$  \\[+1mm]
  $\Omega_c h^2$                 & $0.1184 \pm 0.0014$         & $0.1183^{+0.0013}_{-0.0015}$        &  $0.11762 \pm 0.00090$    &  $0.11761 \pm 0.00089$  \\[+1mm]
  $H_0$         & $67.87 \pm 0.65$       & $67.91^{+0.67}_{-0.60}$      &  $68.21 \pm 0.41$  &  $68.22 \pm 0.41$  \\[+1mm]
  $\tau$                                             & $0.0571\pm 0.0063$         & $0.0572\pm 0.0063$        &  $0.0576\pm 0.0064$    &  $0.0577\pm 0.0063$    \\[+1mm]
  $\ln(10^{10} A_s)$                                             & $3.036 \pm 0.015$         & $3.036\pm 0.015$        &  $3.035 \pm 0.015$    &  $3.036 \pm 0.015$    \\[+1mm]
  $n_s$              & $0.9685 \pm 0.0046$           & $0.9688 \pm 0.0046$          &  $0.9705 \pm 0.0036$      &  $0.9708 \pm 0.0036$      \\[+1mm]
  $A_L$              & $1.036 \pm 0.055$           & $1.041 \pm 0.038$          &  $1.051 \pm 0.050$      &  $1.050 \pm 0.035$      \\[+1mm]
  \hline \\[-1mm]
  $\Omega_m$                                     & $0.3069^{+0.0080}_{-0.0093}$     & $0.3063^{+0.0077}_{-0.0092}$    &  $0.3023^{+0.0050}_{-0.0057}$     &  $0.3022^{+0.0050}_{-0.0056}$     \\[+1mm]  
  $\sigma_8$                                     & $0.8034\pm 0.0076$     & $0.8032\pm 0.0076$    &  $0.8010\pm 0.0068$     &  $0.8012 \pm 0.0068$     \\[+1mm]    
  $r_d$                               & $147.63\pm 0.28$ & $147.65\pm 0.28$ &  $147.77\pm 0.21$&  $147.76\pm 0.21$     \\[+1mm]  
  \hline\\[-1mm]
  $\chi^2_\text{min}$                               & $30548.57$ & $30556.30$ &  $31968.25$&  $31976.06$     \\[+1mm] 
  $\Delta\chi^2_\text{min}$                               & $+2.53$ & $+3.08$ &  $+1.34$&  $+2.10$     \\[+1mm] 
  $\text{DIC}$                                & $30602.15$ & $30610.84$ &  $32022.17$&  $32028.49$     \\[+1mm]  
  $p_D$                              & $26.79$ & $27.27$ &  $26.88$&  $26.22$     \\[+1mm]
  $\Delta\text{DIC}$                                & $-2.51$ & $-1.86$ &  $-0.30$&  $+0.92$     \\[+1mm]
\end{tabular}
\\[+1mm]
\end{adjustbox}
\caption{{\bf $\Lambda$CDM + $A_L$}: Mean and 68\% confidence limits of the primary and derived parameters of the $\Lambda$CDM+$A_L$ model obtained after analyzing the datasets: PR4, PR4+lensing, PR4+BAO+SNIa and PR4+lensing+nonCMB. The $H_0$ has units of km/s/Mpc whereas $r_d$ has units of Mpc. We also include the values of $ \chi^2_{\text{min}}$ and DIC together with the differences with respect to the $\Lambda$CDM model, denoted by $\Delta\chi^2_{\text{min}}$ and $\Delta$DIC, respectively. We include $p_D$, the effective number of free parameters.}
\label{tab:results_LCDM_AL}
\end{table*}



\begin{table*}
\begin{adjustbox}{width=1\textwidth}
\small
\begin{tabular}{lccccc}

  Parameter                     &  BAO+SNIa                     & PR4                         &  PR4+lensing               &  PR4+BAO+SNIa        & PR4+lensing+BAO+SNIa     \\[+1mm]
 \hline \\[-1mm]
  $\Omega_b h^2$                & $0.0266^{+0.012}_{-0.0050}$          & $0.02227\pm 0.00014$       & $0.02226\pm 0.00013$      &  $0.02232\pm 0.00012$  &  $0.02231 \pm 0.00012$  \\[+1mm]
  $\Omega_c h^2$                & $0.121\pm 0.024$  & $0.1186\pm 0.0012$         & $0.1187\pm 0.0011$        &  $0.11753 \pm 0.00097$    &  $0.11791 \pm 0.00092$  \\[+1mm]
  $H_0$       & $70^{+9}_{-5}$  & $85^{+10}_{-6}$       & $85^{+10}_{-6}$     &  $67.55 \pm 0.70$  &  $67.58 \pm 0.70$  \\[+1mm]
  $\tau$                        & $0.0576$                     & $0.0576\pm 0.0063$         & $0.0579\pm 0.0061$        &  $0.0597\pm 0.0064$    &  $0.0614\pm 0.0062$    \\[+1mm]
  $\ln(10^{10} A_s)$                         & $2.75\pm 0.63$                     & $3.038\pm 0.014$         & $3.039\pm 0.012$        &  $3.041 \pm 0.014$    &  $3.050 \pm 0.012$    \\[+1mm]
  $n_s$            & $0.9678$   & $0.9678\pm 0.0041$           & $0.9677\pm 0.0040$          &  $0.9704 \pm 0.0037$      &  $0.9698 \pm 0.0036$      \\[+1mm]
  $w_0$            & $-0.925\pm 0.049$   & $-1.51^{+0.18}_{-0.35}$           & $-1.52^{+0.18}_{-0.33}$          &  $-0.978 \pm 0.026$      &  $-0.984 \pm 0.026$      \\[+1mm]
  \hline \\[-1mm]
  $\Omega_m$                           & $0.299\pm 0.014$           & $0.204^{+0.017}_{-0.060}$     & $0.202^{+0.017}_{-0.058}$    &  $0.3080\pm 0.0068$     &  $0.3086 \pm 0.0067$     \\[+1mm]  
 $\sigma_8$                           & $0.70^{+0.14}_{-0.31}$           & $0.946^{+0.095}_{-0.050}$     & $0.950^{+0.089}_{-0.048}$    &  $0.797\pm 0.011$     &  $0.8031 \pm 0.0092$     \\[+1mm]  
  $r_d$                         & $144.4^{+7.6}_{-1.9}$       & $147.60\pm 0.26$ & $147.58\pm 0.25$ &  $147.81\pm 0.23$&  $147.72\pm 0.21$     \\[+1mm]  
  \hline\\[-1mm]
  $\chi^2_\text{min}$                         & $1415.40$       & $30549.08$ & $30557.23$ &  $31969.25$&  $31977.40$     \\[+1mm]  
  $\Delta\chi^2_\text{min}$                   & $+2.88$            & $+2.02$ & $+2.15$ &  $+0.34$&  $+0.76$     \\[+1mm]  
  $\text{DIC}$                         & $1421.34$       & $30600.12$ & $30608.87$ &  $32022.13$&  $32033.22$     \\[+1mm]  
  $p_D$                      & $2.97$        & $25.52$ & $25.82$ &  $26.44$&  $27.91$     \\[+1mm] 
  $\Delta\text{DIC}$                         & $+0.96$       & $-0.48$ & $+0.11$ &  $-0.26$&  $-1.58$     \\[+1mm]  
\end{tabular}
\\[+1mm]
\end{adjustbox}
\caption{{\bf $w_0$CDM}: Mean and 68\% confidence limits of the primary and derived parameters of the $w_0$CDM model obtained after analyzing the datasets: BAO+SNIa, PR4, PR4+lensing, PR4+BAO+SNIa and PR4+lensing+nonCMB. The $H_0$ has units of km/s/Mpc whereas $r_d$ has units of Mpc. We also include the values of $ \chi^2_{\text{min}}$ and DIC together with the differences with respect to the $\Lambda$CDM model, denoted by $\Delta\chi^2_{\text{min}}$ and $\Delta$DIC, respectively. We include $p_D$, the effective number of free parameters.}
\label{tab:results_w0CDM}
\end{table*}



\begin{table*}
\begin{adjustbox}{width=1\textwidth}
\small
\begin{tabular}{lccccc}

  Parameter                     &  BAO+SNIa                     & PR4                         &  PR4+lensing               &  PR4+BAO+SNIa        & PR4+lensing+BAO+SNIa     \\[+1mm]
 \hline \\[-1mm]
  $\Omega_b h^2$                & $0.0270^{+0.012}_{-0.0047}$          & $0.02227\pm 0.00014$       & $0.02227\pm 0.00014$      &  $0.02225\pm 0.00013$  &  $0.02224 \pm 0.00013$  \\[+1mm]
  $\Omega_c h^2$                & $0.135\pm 0.032$  & $0.1185\pm 0.0012$         & $0.1185\pm 0.0012$        &  $0.1186 \pm 0.0011$    &  $0.11881 \pm 0.00098$  \\[+1mm]
  $H_0$       & $72^{+9}_{-6}$  & $84^{+10}_{-6}$       & $84^{+10}_{-6}$      &  $67.84 \pm 0.72$  &  $67.89 \pm 0.72$  \\[+1mm]
  $\tau$                        & $0.0575$                     & $0.0575\pm 0.0062$         & $0.0576\pm 0.0061$        &  $0.0580\pm 0.0063$    &  $0.0587\pm 0.0061$    \\[+1mm]
  $\ln(10^{10} A_s)$                         & $2.76\pm 0.63$                     & $3.037\pm 0.014$         & $3.038\pm 0.012$        &  $3.039 \pm 0.014$    &  $3.043 \pm 0.012$    \\[+1mm]
  $n_s$            & $0.9681$   & $0.9681\pm 0.0041$           & $0.9680\pm 0.0040$          &  $0.9678 \pm 0.0038$      &  $0.9674 \pm 0.0037$      \\[+1mm]
  $w_0$            & $-0.867^{+0.069}_{-0.078}$   & $-1.27^{+0.41}_{-0.47}$           & $-1.28\pm 0.44$          &  $-0.849 \pm 0.062$      &  $-0.843 \pm 0.062$      \\[+1mm]
  $w_a$            & $-0.53\pm 0.58$   & $-1.03^{+0.74}_{-1.8}$           & $-1.02^{+0.72}_{-1.80}$          &  $-0.59^{+0.28}_{-0.25}$      &  $-0.64^{+0.29}_{-0.24}$      \\[+1mm]
  \hline \\[-1mm]
  $\Omega_m$                           & $0.310^{+0.023}_{-0.015}$           & $0.209^{+0.016}_{-0.065}$     & $0.208^{+0.017}_{-0.065}$    &  $0.3075\pm 0.0069$     &  $0.3076 \pm 0.0068$     \\[+1mm]  
 $\sigma_8$                           & $0.78^{+0.19}_{-0.35}$           & $0.943^{+0.10}_{-0.049}$     & $0.945^{+0.10}_{-0.047}$    &  $0.809\pm 0.012$     &  $0.8131 \pm 0.0098$     \\[+1mm]  
  $r_d$                         & $140.9^{+8.2}_{-20}$       & $147.61\pm 0.26$ & $147.60\pm 0.25$ &  $147.60\pm 0.24$&  $147.56\pm 0.22$     \\[+1mm]  
  \hline\\[-1mm]
  $\chi^2_\text{min}$                         & $1414.78$       & $30547.32$ & $30557.44$ &  $31964.42$&  $31972.85$     \\[+1mm]  
  $\Delta\chi^2_\text{min}$                   & $+3.50$            & $+3.78$ & $+1.94$ &  $+5.17$&  $+5.31$     \\[+1mm]  
  $\text{DIC}$                         & $1422.62$       & $30601.84$ & $30608.62$ &  $32019.50$&  $32028.65$     \\[+1mm]  
  $p_D$                      & $3.92$        & $27.26$ & $25.59$ &  $27.54$&  $27.90$     \\[+1mm] 
  $\Delta\text{DIC}$                         & $-0.32$       & $-2.20$ & $+0.36$ &  $+2.37$&  $+2.99$     \\[+1mm]  
\end{tabular}
\\[+1mm]
\end{adjustbox}
\caption{{\bf $w_0w_a$CDM}: Mean and 68\% confidence limits of the primary and derived parameters of the $w_0w_a$CDM model obtained after analyzing the datasets: BAO+SNIa, PR4, PR4+lensing, PR4+BAO+SNIa and PR4+lensing+nonCMB. The $H_0$ has units of km/s/Mpc whereas $r_d$ has units of Mpc. We also include the values of $ \chi^2_{\text{min}}$ and DIC together with the differences with respect to the $\Lambda$CDM model, denoted by $\Delta\chi^2_{\text{min}}$ and $\Delta$DIC, respectively. We include $p_D$, the effective number of free parameters.}
\label{tab:results_w0waCDM}
\end{table*}

\section*{Acknowledgements}
We would like to thank the insightful discussions with Guadalupe Ca\text{\~{n}}as Herrera, Will Handley, Jes\'us Torrado and we would like to thank David Fern\'andez Sanz for the help setting up the UCM computational facilities.

J. A. acknowledges the support of Universidad Complutense de Madrid UCM project PR3/23-30808, Spanish MINECO/FEDER Grants No. PGC2022-126078NB-
C21 funded by MCIN/AEI/10.13039/ 50110001103 and DGA-FSE grant 2023-E21-23R.
JdCP's research was financially supported by the project "Plan Complementario de I+D+i en el \'area de Astrof{\'\i}sica" funded by the European Union within the framework of the Recovery, Transformation and Resilience Plan - NextGenerationEU and by the Regional Government of Andaluc{\'i}a (Reference AST22\_00001).
Both J. A and J. dCP acknowledge partial support from MICINN (Spain) project PID2022-138263NB-I0 (AEI/FEDER, UE).
Both authors acknowledge participation in the Cost Association Action CA21136 ``Addressing observational
tensions in cosmology with systematics and fundamental physics (CosmoVerse)''
This project has been done by using the resources of the \texttt{Aljuarismi} computing facility of the Facultad de Ciencias F\'isicas of Universidad Complutense de Madrid. We acknowledge the use of the computational resources of CESAR at BIFI Institute (University of Zaragoza), in particular, the facility \texttt{Agustina}.
We wish to ackowledge the use of the \texttt{Cobaya} library \cite{2019ascl.soft10019T,Torrado:2020dgo} for the production of the MCMC chains.


\bibliographystyle{elsarticle-num-names}
\bibliography{references}
\end{document}